\documentclass[journal,10pt]{IEEEtran}
\usepackage{etoolbox}
\usepackage{dtsc-creafig}
\usepackage{notation}

\usepackage{epsfig}
\usepackage{color}
\usepackage{amsmath}
\usepackage{amssymb}
\usepackage{mathabx}
\usepackage{enumerate}
\usepackage{multirow}
\usepackage{bbm}
\usepackage{algorithm}
\usepackage{algorithmic}
\usepackage{lipsum,amsmath,multicol}
\usepackage{stfloats}
\usepackage{arydshln} 
\usepackage{amsfonts}
\usepackage{dsfont}
\usepackage{upgreek}
\usepackage{pgfplots} 
\usepackage{caption}
\usepackage{subfigure} 
\usepackage{booktabs} 

\usepackage{cite}

\definecolor{mycolor}{cmyk}{1,0,1,0}
\definecolor{mycolor1}{rgb}{1.00000,0.50000,0.00000}%
\definecolor{mycolor2}{rgb}{0.00000,0.80000,1.00000}%
\definecolor{mycolor3}{rgb}{1.00000,0.00000,1.00000}%
\definecolor{mycolor4}{rgb}{0.45, 0.31, 0.59}%
\definecolor{mycolor5}{rgb}{0.6, 0.4, 0.8}
\definecolor{carnationpink}{rgb}{1.0, 0.65, 0.79}
\definecolor{auburn}{rgb}{0.43, 0.21, 0.1}

\usepackage{acronym}
\acrodef{LMMSE}{linear minimum mean square error}
\acrodef{FIR}{finite impulse response}
\acrodef{MIMO}{multiple-input multiple-output}
\acrodef{i.i.d.}{independent and identically distributed} 
\acrodef{S/P}{serial to paralell} 
\acrodef{BP}{belief propagation}
\acrodef{SPA}{sum-product algorithm}     
\acrodef{GMP}{Gaussian message passing} 
\acrodef{KL}{Kullback-Leibler} 
\acrodef{ML}{maximum likelihood}
\acrodef{pdf}{probability density function}
\acrodef{SISO}{single-input single-output}
\acrodef{DFE}{decision feedback equalization}
\acrodef{EXIT}{extrinsic information transfer}
\acrodef{BER}{bit error rate}
\acrodef{SER}{symbol error rate}
\acrodef{ECC}{error correction code}
\acrodef{FEC}{forward error correction}
\acrodef{MSE}{mean-square-error}
\acrodef{MMSE}{minimum mean square error}
\acrodef{TLMMSE}{turbo linear minimum mean square error}
\acrodef{APP}{a posteriori probabilities}
\acrodef{LLRs}{log-likelihood ratios}
\acrodef{LLR}{log-likelihood ratio}
\acrodef{BEP}{block expectation propagation}
\acrodef{nuBEP}{non-uniform block expectation propagation}
\acrodef{TBEP}{turbo block expectation propagation}
\acrodef{ISI}{intersymbol interference}
\acrodef{ICI}{intercarrier interference}
\acrodef{AWGN}{additive white Gaussian noise}
\acrodef{MAP}{maximum a posteriori}
\acrodef{LDPC}{low-density parity-check}
\acrodef{EP}{expectation propagation}
\acrodef{SD}{sphere decoding}
\acrodef{MCMC}{Markov chain Monte Carlo}
\acrodef{GTA}{Gaussian tree approximation}
\acrodef{CHEMP}{channel hardening-exploiting message passing}
\acrodef{BP}{belief propagation}
\acrodef{LTI}{linear time invariant}
\acrodef{pmf}{probability mass functions}
\acrodef{KL}{Kullback-Leibler}
\acrodef{KS}{Kalman smoothing}
\acrodef{KSEP}{Kalman smoothing expectation propagation}
\acrodef{SEP}{smoothing expectation propagation}
\acrodef{BP-EP}{belief propagation expectation propagation}
\acrodef{WF}{Wiener filter-type}

\ifCLASSINFOpdf
\else
\fi
\begin{document}
%
\title{A double EP-based proposal for turbo equalization}
%
%
%

\author{Irene~Santos, 
        Juan~Jos\'e~Murillo-Fuentes and Eva~Arias-de-Reyna
\thanks{I. Santos, J.J. Murillo-Fuentes and E. Arias-de-Reyna are with the Dept. Teor\'ia de la Se\~nal y Comunicaciones, Universidad de Sevilla, Camino de los Descubrimientos s/n, 41092 Sevilla, Spain. E-mail: {\tt \{irenesantos,murillo,earias\}@us.es}}
\thanks{This work was partially funded by Spanish government (Ministerio de Econom\'ia y Competitividad TEC2016-78434-C3-R) and by the European Union (FEDER).}}

\maketitle

\begin{abstract}

This letter deals with the application of the expectation propagation (EP) algorithm to turbo equalization. The EP has been successfully applied  to obtain either a better approximation at the output of the equalizer or at the output of the channel decoder to better initialize the Gaussian prior used by the equalizer. In this letter we combine both trends to propose a novel double EP-based equalizer that is able to decrease the number of iterations needed, reducing the computational complexity to twice that of the linear MMSE. This novel equalizer is developed in three different implementations: a block design that exploits  the whole vector of observations, a Wiener filter-type approach that just uses the observations within a predefined window and a Kalman smoothing filter-type approach that emulates  the BCJR behavior. Finally, we include some experimental results to compare the three different implementations and to detail their improvements with respect to other EP-based proposals in the literature. 


\end{abstract}


%
\IEEEpeerreviewmaketitle

\section{Introduction}


Current digital transmissions are corrupted by \ac{ISI} introduced by the dispersive nature of the channels, which negatively affects the received signal. This corrupted signal is processed by the equalizer, that provides an estimation of the transmitted symbols \cite{Haykin09}. These estimations can be probabilistic, resulting in a high benefit for modern channel decoders. In addition, the equalizer and channel decoder can exchange information to improve the estimation, which is known as turbo equalization \cite{Tuchler11}. 

One optimal solution used in turbo equalization is the BCJR \cite{Bahl74}, that obtains the \ac{MAP} probabilities for each transmitted symbol. However, its computational cost
%
increases exponentially with the number of symbols of the constellation and/or the length of the channel, becoming intractable for large channels or high-order modulations. 
In this situation, some approximate inference techniques, such as the \ac{LMMSE}, are employed. 

In turbo equalization, the LMMSE obtains a tractable Gaussian approximation for the \ac{APP} by assuming a Gaussian distribution for the prior according to the channel decoder output. It can be developed in a  block \cite{Muranov10}, \ac{WF} \cite{Koetter02} and \ac{KS} \cite{Park15} implementations. 
However, its performance is far from optimal. 
The \ac{EP} \cite{Minka01,Seeger05} algorithm is a Bayesian inference technique that has been recently applied to turbo equalization to improve the LMMSE performance in its block \cite{Santos16,Santos18}, \ac{WF} \cite{Santos18} and \ac{KS} \cite{Santos17,Santos18c,Sun15,Santos18thesis} implementations. 
The EP shares the structure of the LMMSE, where the estimated Gaussian priors depend on the observations, hence the equalizer is non linear. 

In \cite{Sun15} the EP is developed from a message passing point of view to better approximate with Gaussians the discrete outputs of the channel decoder. To avoid negative variances, they are set to their absolute values. However, this equalizer does not improve the equalization step by itself since it boils down to the LMMSE for standalone equalization, i.e.,  if no turbo equalization is carried out. In contrast, in \cite{Santos16,Santos17,Santos18,Santos18c,Santos18thesis} the output of the decoder is directly projected into the family of Gaussians, as the turbo LMMSE does, i.e., the EP is not applied at the output of the decoder. Instead, the EP is used to obtain a better Gaussian approximation for the extrinsic distribution at the output of the equalizer, before sending it to the channel decoder. This equalizer improves the performance of the LMMSE and the EP in \cite{Sun15}, either as standalone or turbo equalization. In other words, in \cite{Santos16,Santos17,Santos18,Santos18c,Santos18thesis} the EP is used at an inner loop while in \cite{Sun15} it is used at an outer loop, as will be explained in \SEC{EP}. 



In this letter we take advantage of both trends and combine them into a novel double EP-based equalizer. It applies the EP algorithm twice, within the inner and outer loops, outperforming the LMMSE either as standalone or turbo equalization. Also, we improve  the control of negative variances proposed in \cite{Sun15} at the outer loop by setting them to the moments of the information at the output of the channel decoder in case of negative values. In terms of \ac{BER}, this novel equalizer outperforms previous EP-based proposals. Furthermore, it exhibits a significant reduction in complexity. 
Finally, we include some experimental results to show the improvements of this novel double EP equalizer with respect to the others found in the literature \cite{Santos16,Santos17,Santos18,Santos18c,Sun15}. These experiments also include a comparison between the three different implementations (block, \ac{WF} and \ac{KS}) of an EP-based equalizer after averaging over different random channels. 








%
%
%
%



\section{System Model}

A sequence of information bits, $\vect{a}=[\allvect{a}{1}{\k}]\trs$ where $a_i\in\{0,1\}$, is encoded into the codeword  $\vect{b}$. This codeword is partitioned into $\tamframe$ blocks of length $Q=\log_2(\modsize)$ as $\vect{b}=[\vect{b}_{1}, ..., \vect{b}_{\tamframe}]\trs$, where $\vect{b}_{\iter}=[b_{\iter,1}, ..., b_{\iter,Q}]$. Each block $\vect{b}_{\iter}$ is then modulated into a symbol that belongs to a complex $\modsize$-ary constellation with alphabet $\mathcal{A}$ and mean transmitted symbol energy $\energy$. This yields the vector of symbols $\vect{\beforechannel}=[\allvect{\beforechannel}{1}{{\tamframe}}]\trs$ that is transmitted over a channel $\vect{h}=[\allvect{h}{1}{\ntaps}]$ and it is corrupted with \ac{AWGN} whose variance, $\sigma_\noise^2$, is known. The received signal is given by
\begin{align}\LABEQ{smat}
\underbrace{\begin{bmatrix}
\afterchannel_{1} \\ \\ \vdots \\ \\ \afterchannel_{\tamframe+\ntaps-1} 
\end{bmatrix}}_{\vect{\afterchannel}}
=
\underbrace{\begin{bmatrix}
h_1 & & & \matr{0}\\
\vdots & \ddots &  \\
h_{\ntaps} & \ddots & & h_1\\
 & \ddots & & \vdots\\
 \matr{0} & & & h_\ntaps
\end{bmatrix}}_{\matr{H}}
\underbrace{\begin{bmatrix}
\beforechannel_1 \\ \\ \vdots \\ \\ \beforechannel_\tamframe 
\end{bmatrix}}_{\vect{\beforechannel}}
+
\underbrace{\begin{bmatrix}
\noise_1 \\ \\ \vdots \\ \\ \noise_{\tamframe+\ntaps-1}
\end{bmatrix}}_{\vect{\noise}}
\end{align}
where ${\noise_\iter}\sim\cgauss{{\noise_\iter}}{{0}}{\sigma_\noise^2}$.  
%
%
This signal is received by the equalizer that estimates the posterior probability of the transmitted symbol vector as
\begin{equation}\LABEQ{pugiveny}
p(\vect{\beforechannel}|\vect{\afterchannel},\matr{H})\; \propto \;{p(\vect{\afterchannel}|\vect{\beforechannel},\matr{H})\prod\limits_{\iter=1}^{\tamframe} p(\beforechannel_\iter)} 
\end{equation}
where $p(\vect{\afterchannel}|\vect{\beforechannel},\matr{H})=\cgauss{\vect{\afterchannel}}{\matr{H}\vect{\beforechannel}}{\std{\noise}^{2}{\matr{I}}}$ is the likelihood and $p(\beforechannel_\iter)$ is the information on the priors. If the output of the channel decoder is not available, we may assume equiprobable symbols, which is equivalent to setting the prior to a uniform distribution. If available, the information at the output of the channel decoder is fed back to the equalizer and used to update the priors. Note that the computational complexity of \EQ{pugiveny} is proportional to $\modsize^\ntaps$ due to the discrete nature of the symbols. 


In turbo equalization, the equalizer and channel decoder usually exchange extrinsic information. Hence, the equalizer outputs an extrinsic distribution that it is then demapped and given to the channel decoder as extrinsic log-likelihood ratios (LLRs), $L_E(b_{\iter,j})$. 

\section{Double EP turbo equalizer}\LABSEC{EP}

The \ac{EP} algorithm is a Bayesian framework used to approximate a non-tractable distribution, such as \EQ{pugiveny}, with exponential distributions. Recently research works successfully apply this tool to develop a turbo equalizer \cite{Santos16,Santos17,Santos18,Santos18c,Sun15,Santos18thesis}. In \cite{Santos16,Santos17,Santos18,Santos18c}, the EP is used to better approximate the posterior (or extrinsic) distribution at the output of the equalizer. To that end, the non Gaussian factors in \EQ{pugiveny} are replaced by Gaussians, denoted as 
\begin{align}\LABEQ{aproxprior}
t_\iter^{[\ell]}(\beforechannel_\iter) = \cgauss{\beforechannel_\iter}{\mu_{t_\iter}^{[\ell]}}{\sigma_{t_\iter}^{2[\ell]}}, 
\end{align}
yielding the following Gaussian posterior distribution 
\begin{align}\LABEQ{aproxindicator}
\aproxfunc^{[\ell]}(\vect{\beforechannel}) &=\cgauss{\vect{\afterchannel}}{\matr{H}\vect{\beforechannel}}{\std{\noise}^{2}\matr{I}}{\prod\limits_{\iter=1}^{\tamframe} t_\iter^{[\ell]}(\beforechannel_\iter)}. 
\end{align}
The marginal of \EQ{aproxindicator} yields another Gaussian distribution, that we will denote as
\begin{equation}\LABEQ{posterior}
q^{[\ell]}(\beforechannel_\iter)\sim \cgauss{\beforechannel_\iter}{\mu_\iter^{[\ell]}}{\sigma_\iter^{2[\ell]} }
\end{equation}
where {\cite{Santos18}}
\begin{align}
\mu_\iter^{[\ell]}&=\mu_{t_\iter}^{[\ell]}+\sigma_{t_\iter}^{2[\ell]}\vect{h}_{\vect{\iter}}\her \left(\sigma_\noise^2\matr{I}+\matr{H}\boldsymbol{\Sigma}_{t}^{[\ell]}\matr{H}\her\right)\inv (\vect{\afterchannel}-\matr{H}\boldsymbol{\upmu}_{t}^{[\ell]}), \LABEQ{posmeanep} \\
\sigma_\iter^{2[\ell]} & = \sigma_{t_\iter}^{2[\ell]}-\sigma_{t_\iter}^{4[\ell]}\vect{h}_{\vect{\iter}}\her \left(\sigma_\noise^2\matr{I}+\matr{H}\boldsymbol{\Sigma}_{t}^{[\ell]}\matr{H}\her\right)\inv \vecti{h}{\iter},  \LABEQ{posvarep}
\end{align}
$\vect{h}_{\vect{\iter}}$ is the $\iter$-th column of the matrix $\matr{H}$ defined in \EQ{smat}, {${\boldsymbol{\Sigma}_{t}^{[\ell]}}=\mathrm{diag}([\sigma_{t_1}^{2[\ell]}, \hdots, \sigma_{t_{\tamframe}}^{2[\ell]}])$} and $\boldsymbol{\upmu}_{t}^{[\ell]}=[\mu_{t_1}^{[\ell]}, \hdots, \mu_{t_{\tamframe}}^{[\ell]}]\trs$. 

From \EQ{posterior}, one can also compute the extrinsic distribution as
\begin{align}\LABEQ{extrinsic}
q_E^{[\ell]}(\beforechannel_\iter)= q^{[\ell]}(\beforechannel_\iter)/t_\iter^{[\ell]}({\beforechannel_\iter})=
\cgauss{\beforechannel_\iter}{\mu_{E_\iter}^{[\ell]}}{\sigma_{E_\iter}^{2[\ell]}}
\end{align} 
where
\begin{align}\LABEQ{mean_ext}
{\mu_{E_\iter}^{[\ell]}}=\frac{\mu_\iter^{[\ell]}\sigma_{t_\iter}^{2[\ell]}-\mu_{t_\iter}^{[\ell]}\sigma_\iter^{2[\ell]}}{\sigma_{t_\iter}^{2[\ell]}-\sigma_\iter^{2[\ell]}}, \;\;\;\;{\sigma_{E_\iter}^{2[\ell]}}=\frac{\sigma_\iter^{2[\ell]}\sigma_{t_\iter}^{2[\ell]}}{\sigma_{t_\iter}^{2[\ell]}-\sigma_\iter^{2[\ell]}}.
\end{align}

The moments of the Gaussian factors in \EQ{aproxindicator} are updated in parallel and iteratively by matching the moments of the discrete posterior, 
\begin{align}\LABEQ{discreteposterior}
\widehat{p}^{[\ell]}(\beforechannel_\iter)\;\propto\;q_E^{[\ell]}(\beforechannel_\iter)p(\beforechannel_\iter),
\end{align}
and the approximated one, $q_E^{[\ell]}(\beforechannel_\iter)t_\iter^{[\ell+1]}(\beforechannel_\iter)$. A detailed explanation of this procedure is described in \ALG{MMD}. It also includes a damping factor ($\beta$), a minimum allowed variance ($\epsilon$) and a control of negative variances to improve convergence and control instabilities. 
This procedure is repeated iteratively and we will refer to it as \textit{inner} loop. 

\begin{algorithm}[!tb]
\begin{algorithmic}
\STATE 
{\bf Given inputs}: $p(\beforechannel_\iter)$, $t_\iter^{[\ell]}(\beforechannel_\iter)$ with moments $\mu_{t_\iter}^{[\ell]},\sigma_{t_\iter}^{2[\ell]}$ and $q_E^{[\ell]}(\beforechannel_\iter)$ with moments $\mu_{E_\iter}^{[\ell]},\sigma_{E_\iter}^{2[\ell]}$ 
\STATE
\vspace{0.2cm}
1) Compute the moments $\mu_{{\widehat{p}}_\iter}^{[\ell]},\sigma_{{\widehat{p}}_{\iter,aux}}^{2[\ell]}$ of the discrete posterior, $\widehat{p}^{[\ell]}(\beforechannel_\iter)$, defined in \EQ{discreteposterior}. 
Set the variance value taking into account a {\it minimum allowed variance},
\mbox{$\sigma_{{\widehat{p}}_\iter}^{2[\ell]}=\max(\epsilon,\sigma_{{\widehat{p}}_{\iter,aux}}^{2[\ell]})$}. 
\STATE
2)  Run \textit{moment matching}:  Set the mean and variance of the unnormalized Gaussian distribution  $q_E^{[\ell]}(\beforechannel_\iter)\cgauss{\beforechannel_\iter}{\mu_{t_\iter,new}^{[\ell{+1}]}}{\sigma_{t_\iter,new}^{2[\ell{+1}]}}$ 
equal to $\mu_{{\widehat{p}}_\iter}^{[\ell]}$ and $\sigma_{{\widehat{p}}_\iter}^{2[\ell]}$, to get the solution
\begin{align}
\sigma_{t_\iter,new}^{2[\ell+1]}=\frac{\sigma_{{\widehat{p}}_\iter}^{2[\ell]}\sigma_{E_\iter}^{2[\ell]}}{\sigma_{E_\iter}^{2[\ell]}-\sigma_{{\widehat{p}}_\iter}^{2[\ell]}} , \;\;\;
\mu_{t_\iter,new}^{[\ell+1]}= \frac{\mu_{{\widehat{p}}_\iter}^{[\ell]}\sigma_{E_\iter}^{2[\ell]}-{\mu^{[\ell]}_{E_\iter}}\sigma_{{\widehat{p}}_\iter}^{2[\ell]}}{\sigma_{E_\iter}^{2[\ell]}-\sigma_{{\widehat{p}}_\iter}^{2[\ell]}}.
\end{align}
\STATE
3) Run \textit{damping}: Update the values as
\begin{align}
\sigma_{t_\iter}^{2[\ell+1]}&=\left(\beta\frac{1}{\sigma_{t_\iter,new}^{2[\ell+1]}} + (1-\beta)\frac{1}{\sigma_{t_\iter}^{2[\ell]}}\right)\inv \LABEQ{Lambdak1} ,\\
\mu_{t_\iter}^{[\ell+1]}&=\sigma_{t_\iter}^{2[\ell+1]}\left(\beta \frac{\mu_{t_\iter,new}^{[\ell+1]}}{\sigma_{t_\iter,new}^{2[\ell+1]}} + (1-\beta)\frac{\mu_{t_\iter}^{[\ell]}}{\sigma_{t_\iter}^{2[\ell]}}\right). \LABEQ{Lambdak2}
\end{align}
\STATE
4) Control of \textit{negative variances}:
\IF{${\sigma_{t_\iter}^{2[\ell+1]}}<0$}
\vspace{-0.3cm}
\STATE
\begin{align}
\sigma_{t_\iter}^{2[\ell+1]}=\sigma_{t_\iter}^{2[\ell]}, \,\,\,\,\,\,\,\, \mu_{t_\iter}^{[\ell+1]}=\mu_{t_\iter}^{[\ell]}. 
\end{align}
\ENDIF
\STATE
{\bf Output}: $\mu_{t_\iter}^{[\ell+1]},\sigma_{t_\iter}^{2[\ell+1]}$
\end{algorithmic}
\caption{Moment Matching and Damping at iteration $\ell$}\LABALG{MMD}
\end{algorithm}

After $\iterep$ iterations of the previous EP procedure, the extrinsic distributions, $q_E^{[\iterep+1]}(\beforechannel_\iter)$, are sent to the channel decoder, whose output is fed back to the equalizer and used to update the information on the priors, $p(\beforechannel_\iter)$. This procedure is repeated along $\nturbo$ iterations and we will refer to it as \textit{outer} loop. 
Since the information provided by the channel decoder, $p(\beforechannel_\iter)$, is discrete, the first step of the equalizer is to find an initial Gaussian approximation, $t_\iter^{[1]}(\beforechannel_\iter)$. In \cite{Santos16,Santos17,Santos18,Santos18c}, this Gaussian approximation is obtained by projecting $p(\beforechannel_\iter)$ into the family of Gaussians, as the turbo LMMSE does, i.e., 
\begin{align}\LABEQ{priorInitial}
t_\iter^{[1]}(\beforechannel_\iter)=\mbox{Proj}_G[p(\beforechannel_\iter)]\sim \cgauss{\beforechannel_\iter}{\mu_{t_\iter}^{[1]}}{\sigma_{t_\iter}^{2[1]}}
\end{align}
where $\mu_{t_\iter}^{[1]}=\E_p[\beforechannel_\iter]$ and $\sigma_{t_\iter}^{2[1]}=\E_p[(\beforechannel_\iter-\mu_{t_\iter}^{[1]})^2]$.

In this manuscript, we propose a different Gaussian approximation for $p(\beforechannel_\iter)$ 
%
that has the same computational complexity as \EQ{priorInitial} and more accurate results. Specifically, we apply EP at the output of the decoder, as introduced by the proposal in \cite{Sun15}. Since we have already proposed the use of EP in the inner loop, we are proposing a second EP that takes as a starting point the extrinsic distribution that was given to the channel decoder at the previous turbo iteration, $q_E^{[\iterep+1]}(\beforechannel_\iter)$, yielding
%
%
\begin{equation}\LABEQ{EPPEN}
t_\iter^{[1]}(\beforechannel_\iter)=\frac{\mbox{Proj}_G[\widehat{p}^{[\iterep+1]}(\beforechannel_\iter)]}{q_E^{[\iterep+1]}(\beforechannel_\iter)} \sim \cgauss{\beforechannel_\iter}{\mu_{t_\iter}^{[1]}}{\sigma_{t_\iter}^{2[1]}},
\end{equation}
where 
\begin{align}
\mu_{t_\iter}^{[1]}&= 
\frac{\sigma_{E_\iter}^{2[\iterep+1]}\mu^{[\iterep+1]}_{\widehat{p}_\iter}-\sigma_{\widehat{p}_\iter}^{2{[\iterep+1]}}\mu_{E_\iter}^{{[\iterep+1]}}}{\sigma_{E_\iter}^{2[\iterep+1]}-\sigma_{\widehat{p}_\iter}^{2{[\iterep+1]}}}, \LABEQ{meanEP2} \\
\sigma_{t_\iter}^{2[1]}&=\frac{\sigma_{\widehat{p}_\iter}^{2{[\iterep+1]}}{\sigma_{E_\iter}^{2[\iterep+1]}}}{\sigma_{E_\iter}^{2[\iterep+1]}-\sigma_{\widehat{p}_\iter}^{2{[\iterep+1]}}}   \LABEQ{varEP2}
\end{align}
and $(\mu_{\widehat{p}_\iter}^{[\iterep+1]},\sigma_{\widehat{p}_\iter}^{2[\iterep+1]})$ are the moments of $\widehat{p}^{[\iterep+1]}(\beforechannel_\iter)$, defined in \EQ{discreteposterior}. If the variances in \EQ{varEP2} lead to negative values we replace the moments of \EQ{EPPEN} by the results  of the method in \EQ{priorInitial}. 
The whole procedure is detailed in \ALG{DEP}. 

\begin{algorithm}[!tb]
\begin{algorithmic}
\STATE 
{\bf Inputs}: $\vect{h}$, $\sigma_\noise^2$ and $\afterchannel_\iter$ for $\iter=1,...,\tamframe+\ntaps-1$
\STATE
1) Initialization: Set $p(\beforechannel_\iter)=\frac{1}{\modsize}\sum_{\beforechannel\in\mathcal{A}} \delta(\beforechannel_\iter-\beforechannel)$ and $q_E^{[\iterep+1]}(\beforechannel_\iter)=1$ for $\iter=1,...,\tamframe$
%
\FOR {$t=0,...,\nturbo$}
\STATE
\textbf{EP at the \textit{outer} loop:} \\
2) Compute $t_\iter^{[1]}(\beforechannel_\iter)$ as in \EQ{EPPEN} 
and compute its moments $\mu_{t_\iter}^{[1]},\sigma_{t_\iter}^{2[1]}$ 
\STATE
3) Control of negative variances: \\
\IF {$\sigma_{t_\iter}^{2[1]}<0$}
\STATE
Set $t_\iter^{[1]}(\beforechannel_\iter)=\mbox{Proj}_G[p(\beforechannel_\iter)]$ and compute its moments $\mu_{t_\iter}^{[1]},\sigma_{t_\iter}^{2[1]}$
\ENDIF \\
\textbf{EP at the \textit{inner} loop:} \\
\FOR {$\ell=1,...,\iterep$}
\FOR {$\iter=1,...,\tamframe$}
\STATE
4) Compute the $\iter$-th extrinsic distribution, $q_E^{[\ell]}(\beforechannel_\iter)$. 
%
%
\STATE
5) Run \ALG{MMD} with $p(\beforechannel_\iter)$, $t_\iter^{[\ell]}(\beforechannel_\iter)$ and $q_E^{[\ell]}(\beforechannel_\iter)$ to obtain $ \mu_{{t}_\iter}^{[\ell+1]},\sigma_{t_\iter}^{2[\ell+1]}$. \\
\ENDFOR 
\ENDFOR 
\STATE
6) With the values $\mu_{t_\iter}^{[\iterep+1]}$ and $\sigma_{t_\iter}^{2[\iterep+1]}$ computed after EP, calculate the extrinsic distribution $q_E^{[\iterep+1]}(\beforechannel_\iter)$. 
\STATE
7) Demap the extrinsic distribution and compute the extrinsic LLR, $L_E(b_{\iter,\iterj})$. 
\STATE
8) Run the channel decoder to output $p(\beforechannel_\iter)$
\ENDFOR 
\STATE
{\bf Output}: Deliver $L_E(b_{\iter,\iterj})$ to the decoder  for $\iter=1,...,\tamframe$ and $j=1,\hdots,Q$
\end{algorithmic}
\caption{Double EP Turbo Equalizer}\LABALG{DEP}
\end{algorithm}


Note that in this proposal the EP is applied twice, as showed in \FIG{TEPDouble}. We first use EP in an \textit{inner} loop ($\ell=1,..,\iterep$) to obtain a Gaussian extrinsic distribution at the output of the equalizer, $q_E^{[\ell]}(\beforechannel_\iter)$. This use of EP is plotted as a gray block named EP$_1$. Then, a second EP is used within an \textit{outer} loop ($t=0,...,\nturbo$) to  find an initial Gaussian approximation, $t_\iter^{[1]}(\beforechannel_\iter)$, for the discrete information at the output of the channel decoder, see EP$_2$ in \FIG{TEPDouble}. 
At this point, it is important to remark the difference with previous proposals, where the EP is applied just once. In \cite{Santos16,Santos17,Santos18,Santos18c}, the EP$_2$ block is replaced by a projection into a Gaussian distribution, as described in \EQ{priorInitial}. On the other hand, in the proposal in \cite{Sun15}, named BP-EP, the block EP$_1$ does not appear, i.e., BP-EP can be viewed as a particularization of the scheme in \FIG{TEPDouble} where $\iterep$ is set to 0, boiling down to the LMMSE for standalone equalization. Also, the control of negatives variances is different since {BP-EP} takes the absolute values for the negative variances. 

%

\begin{figure}[htb]
\centering
\scalebox{0.6}{\tikzset{block/.style = {draw, rectangle, line width=0.5pt,
minimum height=1cm, minimum width=1.5cm,rounded corners,font=\Large},
input/.style = {coordinate},
output/.style = {coordinate},
virtual/.style={coordinate},
blockcirc/.style = {draw, circle, line width=0.5pt,
minimum height=0.5cm, minimum width=0.5cm,rounded corners},
window/.style={rectangle,draw,dashed,minimum width=5.05cm,minimum height=4.4cm,line width=1.1pt},
window2/.style={rectangle,draw,dashed,minimum width=10.04cm,minimum height=4.2cm,line width=1pt},
blockgray/.style = {draw=gray, fill=gray!30, rectangle, line width=0.5pt,
minimum height=1cm, minimum width=1.5cm,rounded corners,font=\Large},
}

\begin{tikzpicture}[auto, >={Latex[width=5pt,length=5pt]}]
\node [input, name=input] {};  
\node [block, right =1 cm of input, name=eq] {LMMSE};
\node [virtual, right =1.6 cm of eq, name=v1] {};
\node [block, minimum width=1.4cm, right =2.5 cm of v1, name=demap] {Demap};
\node [block, right =1.8 cm of demap, name=decoder] {\begin{tabular}{c}Channel \\ Decoder\end{tabular}};
\path [->] (eq) -- node[name=u2] {} (v1); 
\node [virtual, below = 3.1 cm of u2] (feedback) {};
\node [block, minimum width=1.0cm, below = 1.4 cm of u2, right = 5.5cm of feedback, name=map] {Map};

\draw [->] (input) -- node {\Large$\displaystyle \afterchannel_\iter\hspace{0.2cm}$} (eq);
\draw [-] (eq) -- node {} (v1);
\draw [->] (v1) -- node {\Large$\displaystyle {q_E(\beforechannel_\iter)}\hspace{2.1cm}$} (demap);
\draw [->] (demap) -- node {\Large$\displaystyle L_E(b_{{k,j}})$} (decoder);

\draw [-,color=white] (v1.south) -- node[left,color=black] {\Large$\iterep$ loops $\hspace{-0.25cm}$} (2.9,-2.35);
\draw [-,color=white] (v1.south) -- node[left,color=black] {\Large (inner) $\hspace{-0.25cm}$} (2.9,-3.60);
\draw [-,color=white] (v1.south) -- node[right,color=black] {\Large$\hspace{2.9cm}$ $\nturbo$ loops} (6.70,-2.35);
\draw [-,color=white] (v1.south) -- node[right,color=black] {\Large $\hspace{3.0cm}$ (outer)} (6.70,-3.60);

\node [blockgray, minimum width=1.2cm, left =2.9 cm of map, name=proj] {$\mbox{EP}_2$};
\draw [-] (proj) -- node[pos=0.86,name=sw] {\mysw} node {} (feedback);
\draw [->] (feedback) -| node[right,pos=0.58] {\Large$\hspace{0.3cm}\displaystyle {t_\iter(\beforechannel_\iter)}$}(eq) ;
\node [blockgray, minimum width=1.1cm,below =0.5 cm of v1, name=ep] {$\mbox{EP}_1$};

\node [virtual, below = 1.3 cm of ep] (feedbackEP) {};
\draw [-] (ep.south) -- node  [right,pos=0.35] {} (feedbackEP);

\draw [->] (v1) -- node {} (ep);
\draw [->] (map) -- node[above] {\Large$\hspace{0.3cm}\displaystyle p(\beforechannel_\iter)$} (proj);
\node [output, right = 0.8 cm of decoder] (output) {};
\draw [->] (decoder) -- node {\Large$\displaystyle \widehat{a}_i$}(output);
\draw [->] (decoder.south) |- node[right,pos=0.38] {\Large$\displaystyle L(b_{{k,j}})$}(map);

\filldraw[fill=white,even odd rule](4.6,-2.81) circle(0.05);
\filldraw[fill=white,even odd rule](4.6,-3.1) circle(0.05);

\node[window, name=win1,minimum width=6.1cm,minimum height=4.7cm] at (3.75,-1.5) {};

\end{tikzpicture}}
\caption{\small Turbo Double EP-based receiver diagram} \LABFIG{TEPDouble}
\end{figure}
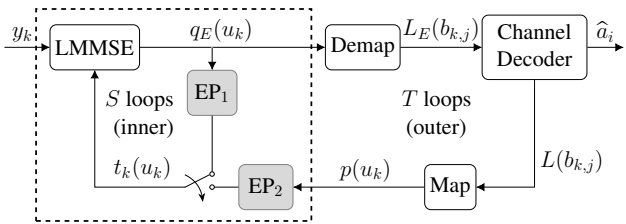


\section{Three different implementations}\LABSEC{implementation}

The computational complexity of \ALG{DEP} is dominated by the computation of the extrinsic distribution, $q_E^{[\ell]}(\beforechannel_\iter)$, at steps 4 and 6. This distribution has to be recomputed along a few EP iterations and turbo iterations, yielding a final complexity of $\order(K\varepsilon_{eq})$, where $K=(\iterep+1)(\nturbo+1)$ and $\varepsilon_{eq}$ denotes the cost of computing $q_E^{[\ell]}(\beforechannel_\iter)$ within the equalizer.

%
%

In its block implementation, the estimation of the transmitted symbols takes into account the whole vector of observations \cite{Santos16,Santos18} and the moments of $q_E^{[\ell]}(\beforechannel_\iter)$ are obtained with \EQ{posmeanep} and \EQ{posvarep}. The inversion in \EQ{posvarep}
can be solved with $\varepsilon_{eq}=\ntaps\tamframe^2$ if the banded-structure of the matrix is exploited \cite{Santos16}. Hence, the final complexity is $\order(K\ntaps\tamframe^2)$. 


 To reduce the computational complexity of the block proposal, a \ac{KS} approach is proposed in \cite{Santos17} {and improved in \cite{Santos18c}}. It merges both forward and backward estimations into a smoothing one, emulating  the BCJR behavior. 
This proposal exhibits the same performance as its block counterpart with linear complexity in the frame length, $\varepsilon_{eq}=\tamframe{\ntaps^2}$. Its final complexity is $\order(K\tamframe{\ntaps^2})$.
%
The closed-form expression for $q_E^{[\ell]}(\beforechannel_\iter)$ to be used at step 4 and 6 of \ALG{DEP} is detailed in eqn. {(3.40) of \cite{Santos18thesis}.} We will denote this proposal as double \ac{KS} \ac{EP} (D-KSEP). As discussed in \SEC{EP}, if $\iterep=0$ and step 3 is replaced by taking the absolute value of $\sigma_{t_k}^{2[1]}$, then it yields the BP-EP \cite{Sun15}. Its computational complexity is $\order({(T+1)}\tamframe\ntaps^2)$. 


Finally, a \ac{WF} approach can be exploited to reduce the computational complexity  to be quadratic in the length of a to-be-predefined window, $\winsize$, i.e.,  $\varepsilon_{eq}=\tamframe\winsize^2$, yielding a complexity $\order(K\tamframe\winsize^2)$ \cite{Santos18}.
%
However, its performance degrades in comparison to the block or \ac{KS} designs since it just uses $\winsize$ observations 
\cite{Santos18thesis}. In this implementation, eqn. {(28)} of \cite{Santos18} is used at steps 4 and 6 of \ALG{DEP} to compute $q_E^{[\ell]}(\beforechannel_\iter)$. We will denote this proposal as double filter \ac{EP} (D-FEP). 

The number of EP iterations, $\iterep$, and of turbo iterations, $\nturbo$, must be set to speed up convergence while minimizing $K$. Convergence is also driven by the rest of EP parameters, $\beta$ and $\epsilon$. In \cite{Santos18}, the EP parameters are optimized to $\epsilon=10^{-8}$, $\beta=\min(\exp^{t/1.5}/10,0.7)$, $\iterep=3$ and $\nturbo=5$.  For the double-EP algorithms proposed, we adopt these values for $\beta$ and $\epsilon$ while the proposed improvement in the estimation of the probabilities at the output of the channel decoder allows to reduce the number of inner iterations to $\iterep=1$.
 
%

\section{Experimental results}\LABSEC{results}


In \FIG{128QAM7taps} we compare the \ac{BER} of several equalizers after the channel decoder for (a) a $64$-QAM and (b) a $128$-QAM. The results are averaged over $100$ random channels and $10^4$ random encoded words of length $\t=4096$ (per channel realization). Each channel tap is zero mean Gaussian independently distributed. The absolute value of LLRs given to the decoder is limited to $5$ to avoid very confident probabilities. A (3,6)-regular \ac{LDPC} of rate $1/2$ is used, for a maximum of $100$ iterations. First, we include in dash-dotted the LMMSE in its block implementation. 
Then we depict in dashed lines the \ac{BER} of previous inner EP approaches with $\iterep=3$ and $\nturbo=5$, i.e. $K=24$: the block EP (BEP) ($\circ$) {\cite{Santos18}}, \ac{WF} EP (FEP) ($\diamond$) \cite{Santos18} and KS EP (KSEP) ($\circ$) \cite{Santos18c}. It has been checked that smaller values of $\iterep$ and/or $\nturbo$ lead to a degradation of \ac{BER}. Note that the BEP and KSEP are depicted with the same marker as they exhibit the same BER  \cite{Santos18c}. Also, we simulate the BP-EP proposal \cite{Sun15} until convergence, resulting $T=8$ ($K=9$) for 64-QAM and $T=11$ ($K=12$) for 128-QAM, in dotted lines. Finally, we include (in solid) the \ac{BER} for the double-EP solutions proposed: D-BEP {({$\circ$})}, D-FEP {($\diamond$)} and D-KSEP {({$\circ$})}. We use $\iterep=1$, $\nturbo=5$, i.e. $K=12$. D-BEP and D-KSEP are depicted as one, since they have the same BER. These ones are also included (as $\square$) for $\nturbo=3$, i.e. $K=8$.
%

%
%



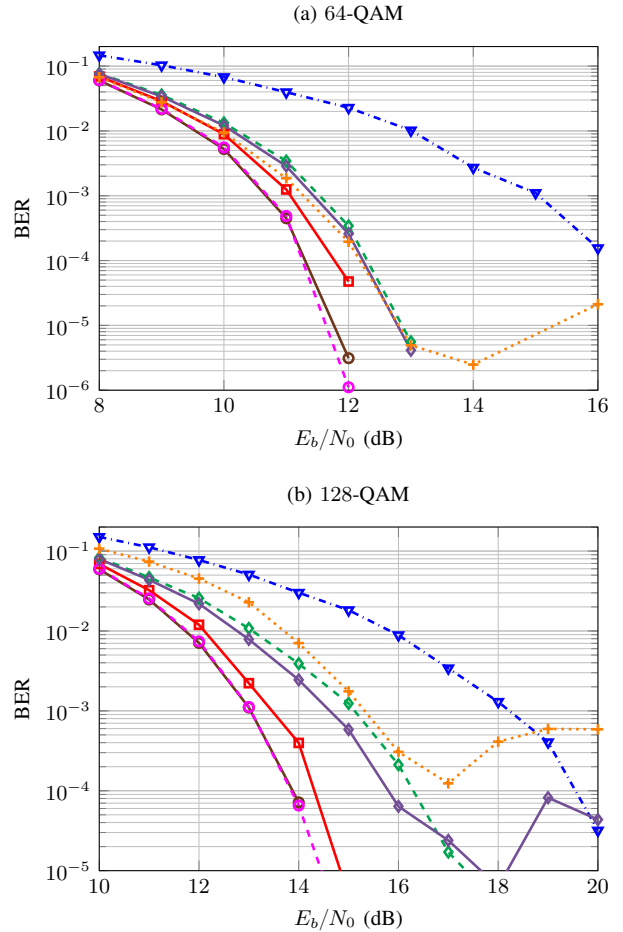
\begin{figure}[htb]
\centering
\scalebox{0.9}{\begin{tikzpicture}[scale=1]

\begin{axis}[%
width=2.9in,
height=2in,
scale only axis,
every axis/.append style={font=\small},
xmajorgrids,
xmin=8,
xmax=16,
xlabel style={align=center}, xlabel={$\EbNo$ (dB)},
ymode=log,
ymin=1e-06,
ymax=2e-1,
yminorticks=true,
ylabel={BER},
name=plot1,
ymajorgrids,
yminorgrids,
title={(a) $64$-QAM},
legend style={at={(1,1)},anchor=north east,draw=black,fill=white,legend cell align=left
,font=\footnotesize
},yshift=-1cm
]

\addplot [color=blue,dashdotted,mark=triangle,mark size=2.2,line width=1.1,mark options={solid,,rotate=180}]
  table[row sep=crcr]{
8	0.146659008189474	\\
9	0.102988026610526	\\
10	0.0676474062736842	\\
11	0.0395697452315789	\\
12	0.022788072368421	\\
13	0.0101980994105263	\\
14	0.00271274694736842	\\
15	0.00110355967368421	\\
16	0.000155432536842105	\\
17	0	\\
18	0	\\
};


\addplot [color=auburn,solid,mark=o,mark size=2,line width=1.1,mark options={solid}]
  table[row sep=crcr]{
8	0.0599917466210526	\\
9	0.0214547725263158	\\
10	0.00524950907368421	\\
11	0.000449136715789474	\\
12	3.11981052631579e-06	\\
13	0	\\
};

\addplot [color=red,solid,mark=square,mark size=1.7,line width=1.1,mark options={solid}]
  table[row sep=crcr]{
8	0.0691620938505263	\\
9	0.0287691700631579	\\
10	0.00890256117368421	\\
11	0.00124937502105263	\\
12	4.75378947368421e-05	\\
13	0	\\
};

\addplot [color=mycolor3,dashed,mark=o,mark size=2,line width=1.1,mark options={solid}]
  table[row sep=crcr]{
8	0.0606533684421053	\\
9	0.0219154082105263	\\
10	0.00555121468421053	\\
11	0.000485989305263158	\\
12	1.11021052631579e-06	\\
13	0	\\
14	0	\\
15	0	\\
16	0	\\
17	0	\\
18	0	\\
};

\addplot [color=mycolor,dashed,mark=diamond,mark size=2.2,line width=1.1,mark options={solid}]
  table[row sep=crcr]{
8	0.0768189301589473	\\
9	0.0358823526347368	\\
10	0.0133068703473684	\\
11	0.00342736823157895	\\
12	0.000342994378947368	\\
13	5.58182105263158e-06	\\
14	0	\\
15	0	\\
16	0	\\
17	0	\\
18	0	\\
};

\addplot [color=mycolor4,solid,mark=diamond,mark size=2.2,line width=1.1,mark options={solid}]
  table[row sep=crcr]{
8	0.0754475086147369	\\
9	0.0340633971	\\
10	0.0121164068	\\
11	0.00286801112926316	\\
12	0.000258578852631579	\\
13	4.13235789473684e-06	\\
};

\addplot [color=mycolor1,dotted,mark=+,mark size=2.2,line width=1.1,mark options={solid}]
  table[row sep=crcr]{
8	0.0672639669052632	\\
9	0.0276210012105263	\\
10	0.00944047444210527	\\
11	0.00185975710526316	\\
12	0.000195194526315789	\\
13	4.93421052631579e-06	\\
14	2.48252631578947e-06	\\
15	0	\\
16	2.11862105263158e-05	\\
};

\end{axis}

\begin{axis}[%
width=2.9in,
height=2in,
scale only axis,
every axis/.append style={font=\small},
xmajorgrids,
xmin=10,
xmax=20,
xlabel style={align=center}, xlabel={$\EbNo$ (dB)},
ymode=log,
ymin=1e-05,
ymax=2e-1,
yminorticks=true,
ylabel={BER},
at=(plot1.below south west), anchor=above north west,
ymajorgrids,
yminorgrids,
title={(b) $128$-QAM},
legend style={at={(1,1)},anchor=north east,draw=black,fill=white,legend cell align=left
,font=\footnotesize
},yshift=-0.3cm
]

\addplot [color=blue,dashdotted,mark=triangle,mark size=2.2,line width=1.1,mark options={solid,,rotate=180}]
  table[row sep=crcr]{
10	0.150395015273684	\\
11	0.111748372768421	\\
12	0.0773340364526316	\\
13	0.0508357117157895	\\
14	0.0301091604663158	\\
15	0.0182413596842105	\\
16	0.00887034654736842	\\
17	0.00341310329473684	\\
18	0.00129885776842105	\\
19	0.000405680147368421	\\
20	3.18204526315789e-05	\\
21	0	\\
};

\addplot [color=auburn,solid,mark=o,mark size=2,line width=1.1,mark options={solid}]
  table[row sep=crcr]{
10	0.0590830440732632	\\
11	0.0247551748421053	\\
12	0.00709270501052632	\\
13	0.00110069522526316	\\
14	7.15514947368421e-05	\\
15	0	\\
16	0	\\
17	0	\\
18	0	\\
19	0	\\
20	0	\\
21	0	\\
22	0	\\
23	0	\\
24	0	\\
};

\addplot [color=red,solid,mark=square,mark size=1.7,line width=1.1,mark options={solid}]
  table[row sep=crcr]{
10	0.0689404707637895	\\
11	0.0323725440842105	\\
12	0.0119500269	\\
13	0.00222060390315789	\\
14	0.000397081136842105	\\
15	4.84172631578947e-06	\\
16	0	\\
17	0	\\
18	0	\\
19	0	\\
20	0	\\
21	0	\\
22	0	\\
23	0	\\
24	0	\\
};

\addplot [color=mycolor3,dashed,mark=o,mark size=2,line width=1.1,mark options={solid}]
  table[row sep=crcr]{
10	0.0599213500515789	\\
11	0.0255061391263158	\\
12	0.00739696141052632	\\
13	0.00111743324210526	\\
14	6.53832842105263e-05	\\
15	1.04851578947368e-06	\\
16	0	\\
17	0	\\
18	0	\\
19	0	\\
20	0	\\
21	0	\\
};

\addplot [color=mycolor,dashed,mark=diamond,mark size=2.2,line width=1.1,mark options={solid}]
  table[row sep=crcr]{
10	0.0823149638989474	\\
11	0.0467652959578947	\\
12	0.0257460252	\\
13	0.0107765763263158	\\
14	0.0039000502	\\
15	0.0012392843368421	\\
16	0.000211363978947368	\\
17	1.70797368421053e-05	\\
18	3.76746315789474e-06	\\
19	1.61390526315789e-06	\\
20	1.58305263157895e-06	\\
21	0	\\
};

\addplot [color=mycolor4,solid,mark=diamond,mark size=2.2,line width=1.1,mark options={solid}]
  table[row sep=crcr]{
10	0.0793934302101053	\\
11	0.0436908842189474	\\
12	0.0219109307052632	\\
13	0.0078483120631579	\\
14	0.00244348505263158	\\
15	0.000583798894736842	\\
16	6.37542526315789e-05	\\
17	2.39410210526316e-05	\\
18	6.30135789473684e-06	\\
19	8.18666421052632e-05	\\
20	4.35339894736842e-05	\\
21	6.0701052631579e-06	\\
22	2.67268421052632e-06	\\
23	6.15750526315789e-06	\\
24	5.20145263157895e-06	\\
};

\addplot [color=mycolor1,dotted,mark=+,mark size=2.2,line width=1.1,mark options={solid}]
  table[row sep=crcr]{
10	0.107254679065263	\\
11	0.0738198237578948	\\
12	0.0452332261052632	\\
13	0.0228784523157895	\\
14	0.00704586073684211	\\
15	0.00175507052631579	\\
16	0.000306481578947368	\\
17	0.000123751157894737	\\
18	0.000411965263157895	\\
19	0.000594479894736842	\\
20	0.000587993684210527	\\
21	0.000571787789473684	\\
};

\end{axis}

\end{tikzpicture}%
}
\caption{\small Averaged BER along $\EbNo$ for turbo LMMSE (\textcolor{blue}{$\triangledown$}), BEP/KSEP \cite{Santos18,Santos18c}  (\textcolor{mycolor3}{$\circ$}), FEP  \cite{Santos18} (\textcolor{mycolor}{$\diamond$}), BP-EP \cite{Sun15} (\textcolor{mycolor1}{$+$}),
D-BEP/D-KSEP with $T=3$ (\textcolor{red}{$\square$}) and $T=5$ (\textcolor{auburn}{$\circ$}) and D-FEP (\textcolor{mycolor4}{$\diamond$}) 
equalizers, with $\ntaps=7$ for (a) $64$-QAM and (b)$128$-QAM modulations. }
\LABFIG{128QAM7taps}
\end{figure} 

It can be observed that previous inner EP methods 
exhibit a remarkable 3-6 dB gain with respect to the block LMMSE. Compared with the BP-EP they have gains up to 3 dB with quite stable convergence. The instabilities of the BP-EP at large $\EbNo$ are the result of a poor control of negative variances and not performing a damping. 
However, the BP-EP has half {or less of} the computational complexity of the KSEP, with $K=24$. 

The novel double EP equalizers achieve the same or better BER than their inner counterparts, reducing by half the number of iterations needed and hence the computational complexity. 
In \FIG{128QAM7taps}.a the D-KSEP, with $K=8$, has a 0.5 dB gain compared with the BP-EP, with $K=9$, while we have a 1 dB gain if we further iterate the D-KSEP ($K=12$). For 128-QAM in \FIG{128QAM7taps}.b, the D-KSEP has 2-3 dB gains compared with the BP-EP at the same number of iterations. Furthermore, the D-KSEP does not exhibit instabilities at high $\EbNo$. 



The D-FEP improves the performance of the FEP but it is far from the one of other approaches. Also, it presents instabilities at large $\EbNo$ and 128-QAM. 
These drawbacks can be mitigated by using longer windows, i.e. a larger number of observations as input, at the cost of increasing its complexity. 

\section{Conclusions}

In this letter we propose a new double EP-based equalizer where the EP algorithm is applied twice. First, it is used to improve the output of the equalizer, even in the case of no feedback from the channel decoder. Then, the EP is applied to the discrete outputs of the channel decoder, providing a more accurate initialization for the priors used by the {turbo} equalizer. This novel approach can be exploited in block, \ac{WF} and \ac{KS} implementations of the equalizer. 
The experimental results included show that the proposed equalizer improves or achieves the same performance of FEP, BEP and KSEP equalizers \cite{Santos18,Santos18c} with half their computational complexity. 
It also outperforms the LMMSE, with just twice its complexity, and other EP-based solutions, such as the BP-EP \cite{Sun15}. 






%
%

\ifCLASSOPTIONcaptionsoff
  \newpage
\fi



%
%
%

\bibliographystyle{IEEEtran}
\bibliography{BibPaperDouble.bib}

%

%
%
%




\end{document}